\documentclass[sigconf]{acmart}

\AtBeginDocument{%
  \providecommand\BibTeX{{%
    \normalfont B\kern-0.5em{\scshape i\kern-0.25em b}\kern-0.8em\TeX}}}

\setcopyright{acmcopyright}
\copyrightyear{2020}
\acmYear{2020}
\acmDOI{}

\acmConference[KDD DS Health Workshop]{KDD Workshop on Applied Data Science for Healthcare}{August 24, 2020}{San Diego, CA}
\acmBooktitle{[KDD Workshop on Applied Data Science for Healthcare,
  August 24, 2020, San Diego, CA}
\acmPrice{}
\acmISBN{}

\usepackage{xspace}
\usepackage{url}
\usepackage{microtype}                 
\PassOptionsToPackage{warn}{textcomp}  
\usepackage{textcomp}                  
\usepackage{mathptmx}                  
\usepackage{times}                     
\usepackage{tabu}                      
\usepackage{booktabs}                  
\usepackage{amsmath}
\usepackage{amssymb}
\usepackage{mathrsfs}
\usepackage[bottom]{footmisc}
\usepackage{graphics}
\usepackage{multirow}
\usepackage{multicol}
\usepackage{color}
\usepackage{transparent}
\usepackage[export]{adjustbox}
\usepackage{dblfloatfix}
\usepackage{xcolor}
\usepackage{subcaption}
\usepackage{enumitem}

\newcommand{\toolname}{DPVis\xspace}

\newcommand{\heatmap}{Feature Matrix\xspace}

\newcommand{\pathwayfall}{Pathway Waterfall\xspace}

\newcommand{\subjectlist}{Subject List View\xspace}
\newcommand{\kernel}{Dual Kernel Densities View\xspace}

\newcommand{\query}{State Sequence Finder\xspace}
\newcommand{\cohort}{Subgroup List\xspace}

\begin{document}

\title{User-driven Analysis of Longitudinal Health Data\\ with Hidden Markov Models for Clinical Insights}


\author{Bum Chul Kwon}
\affiliation{%
  \institution{IBM Research}
  \city{Cambridge}
  \state{MA}
}
\email{bumchul.kwon@us.ibm.com}

\renewcommand{\shortauthors}{Kwon}

\begin{abstract}
A goal of clinical researchers is to understand the progression of a disease through a set of biomarkers.
Researchers often conduct observational studies, where they collect numerous samples from selected subjects throughout multiple years.
Hidden Markov Models (HMMs) can be applied to discover latent states and their transition probabilities over time.
However, it is challenging for clinical researchers to interpret the outcomes and to gain insights about the disease.
Thus, this demo introduces an interactive visualization system called \toolname, which was designed to help researchers to interactively explore HMM outcomes.
The demo provides guidelines of how to implement the clinician-in-the-loop approach for analyzing longitudinal, observational health data with visual analytics.

\end{abstract}


\begin{teaserfigure}
  \includegraphics[width=\textwidth, frame]{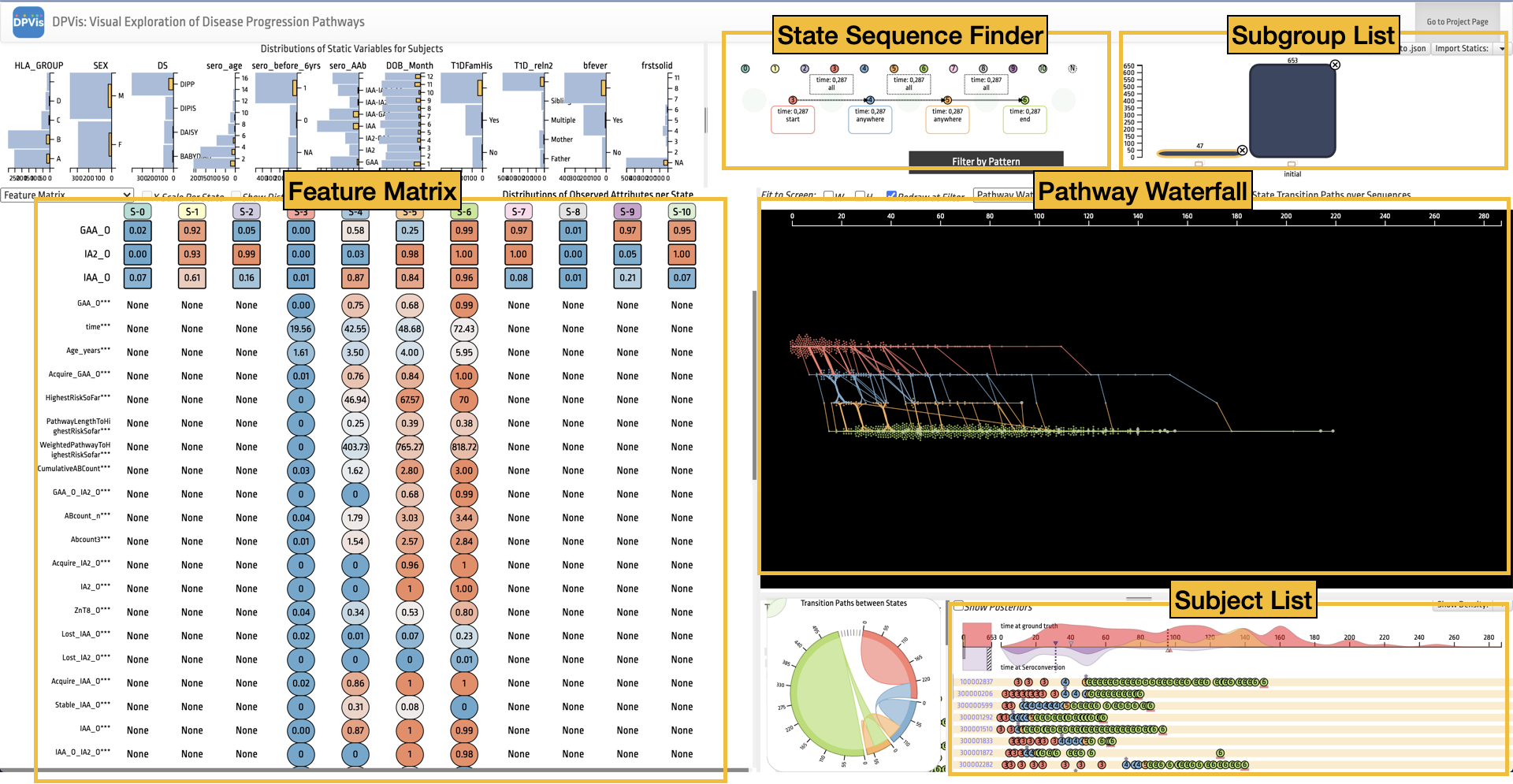}
  \caption{Overview of \toolname. It allows users to gain clinical insights into observational data using various views.}
  \label{fig:teaser}
\end{teaserfigure}

\maketitle

\section{Motivation}

Clinical researchers want to characterize patterns of disease development by using a set of biomarkers.
Therefore, they conduct observational studies to collect samples from selected patients throughout multiple years.  
Large-scale prospective clinical studies make it possible for clinicians to collect serial lab tests, self-reports, and treatments over a long period of time.
The collected data can lead to an understanding of disease progression trajectories of multiple patients.
A major challenge is to find appropriate methods that distill temporal sequences of various measures into clinically useful insights.

One approach for modeling disease progression is to describe a patient's clinical course as a series of transitions between a small number of states, with each state describing a co-occurring pattern of observed measures or symptoms~\cite{wang_unsupervised_2014}. 
The states, commonly referred to as latent or hidden variables, are not directly observed but rather must be discovered from the data.
Hidden Markov models (HMMs) are a class of unsupervised, statistical models that allow for discovering such latent states, understanding characteristics of progression patterns with the associated observed variables for corresponding states, and predicting a patient's future states.
Since HMMs can be trained on longitudinal patient data from large-scale clinical studies and electronic health records (EHRs), HMMs have been successfully applied to capture heterogeneous disease progression patterns in various disease, such as chronic obstructive pulmonary disease~\cite{wang_unsupervised_2014} and Huntington's disease ~\cite{Sun2019_JAMIA}. 

Despite the potential usefulness of HMMs for understanding disease progression patterns, it remains challenging for clinical researchers to make sense of the models and to derive useful insights about the disease. 
Goals of data exploration include discovery of heterogeneous disease progression patterns, observation of interactions between such patterns and patients' characteristics, and testing and refining hypotheses for multiple cohorts in an adhoc manner. 
HMMs can produce relevant outputs that can potentially fulfill these needs, but it is often difficult for clinical researchers to gain useful understanding of clinical observation data by aggregating and connecting various outcomes without the help of interactive visualization systems.
Though prior studies introduced inspiring visualization techniques that can partially solve these problems, no system yet integrates such techniques into a visual analytic solution for disease progression modeling using HMMs.

Thus, this demo introduces a visual analytic system called \toolname \cite{dpvis} that integrates HMM outcomes on observational data with interactive visualizations.
This study demonstrates the usefulness of the visual analytics approach by summarizing design and application of \toolname to observational data from birth-cohort studies of Type 1 Diabetes (T1D).
It provides useful guidelines of how to implement the clinician-in-the-loop approach for gaining clinically meaningful insights into longitudinal, observational health data.

\section{\toolname}

This section describes how the design of \toolname allows clinical researchers to interpret HMMs and to gain clinical insights.
The following sections illustrate how each view allows users to interpret HMM outcomes and to use them for understanding the disease progression patterns.
Full details of DPVis including a video can be found: \url{https://bckwon.com/publication/dpvis}.

The tool is developed using Javascript, HTML, and CSS for the front end and Python for the back end. The team of researchers is actively working on deploying the tool on the web so that other clinical researchers can use the tool. Currently, the tool requires users to train a HMM model on longitudinal data with no personally identifiable information, where each row represents a patient's visit to a clinic. Then, the output includes additional columns, which show an inferred state and posterior probabilities per each visit.
We use a model trained with a type 1 diabetes dataset to illustrate the views of visualizations (see more details of the model in the previous paper~\cite{dpvisamia}.

\subsection{State Summary}

\heatmap summarizes the hidden states of HMMs by showing probabilities of selected variables per state as \autoref{fig:teaser} shows.
The view shows the probabilities of variables in text and intensity of colors with a heatmap.
By using this approach, users can understand the characteristics of states based on the variables while interacting with other views.
For instance, \autoref{fig:teaser} shows probabilities of three islet autoantibodies (IAs) for 11 states discovered by a trained model. 
Researchers can find that States 0, 3, and 8 presents the state without any IAs (low values in blue for all three rows on top).
The three states represent initial states, which indicate that the model discovered three distinctive trajectories starting from each of the three states.
The view also provides probabilities of other variables that are not used in HMMs, which can provide useful insights for steering the model with additional variables.
\heatmap is useful to remind clinical researchers of what the states represent which can be helpful to interpret state transitions in \pathwayfall.

\subsection{State Transition Summary}

\pathwayfall shows subjects' state transition patterns using parallel beeswarm visualizations for subjects' visits overlaid with lines for their trajectories, as \autoref{fig:teaser} shows.
Each dot represents a subject's visit, which is horizontally positioned for the subject's age and vertically positioned for the state discovered for the particular visit.
Trajectory lines are drawn to connect subject's visits, which show the state transition pathways over subjects' age.
Using this view, clinical researchers can view the distribution of visits over time per state.
In addition, they can also gain insights into the heterogeneous state transition patterns over time by viewing bundled paths and following the trajectories.
\autoref{fig:teaser} shows the overview of state transition pathways and the characteristics of states for a particular subcohort, which follows transitions from state 3 through 6 step by step.
According to the view, the majority of transition across four states occur before the age of 60 months.
There are, however, a few outliers who show slow progression, which could be interesting to look at in detail.
In combination with \heatmap, the selected subcohort shows a specific pattern of incrementally adding three IAs, in the order of IAA, IA-2A, and GADA, throughout states 4, 5, and 6, respectively.
\pathwayfall shows an overview of state transition patterns and individual subjects' patterns at the same time.

\subsection{Interactive Cohort Construction}

\query is a way researchers can freely define and refine queries to create a cohort based on state transition patterns.
As Fig.~\ref{fig:teaser} shows, the view includes a canvas, where they can graphically describe state transitions they are looking for.
To increase the customizability, the view also allows them to specify attributes related to states (nodes) and transitions between them (edges).
For example, the current view shows four nodes, namely representing state 4, 5, 6, and 7.
To further specify state transitions, users currently added restrictions on state 4 to be the initial state and state 7 to be the final state, which filters out other subjects.
This view can help clinical researchers to test their hypotheses about state transitions by constructing and comparing multiple subcohorts.
The subcohorts can be available for clinicians in \cohort.

\subsection{Individual Subject Inspection}

\subjectlist allows users to see details of subjects' visits over time and to show summary of health outcomes for selected subjects as \autoref{fig:teaser} shows.
The view shows \kernel, which provides probability density diagrams for selected health outcomes.
For example, clinical researchers chose two health outcomes: seroconversion (purple, below) and onset (red, above).
These two density diagrams show the distribution of age when they show symptoms and get diagnosed at the same time.
This view is particularly useful to evaluate whether specific subcohorts show fast or slow disease progression.
\subjectlist also provides a list of subjects' observations as dots (visits) over a timeline.
Each dot represents a visit, which are color-coded based on the corresponding state.
This view also shows posterior probabilities of state assignment for every visit, which can show additional details of HMM outcomes.
This view allows users to see heterogeneity of state transition patterns by inspecting individual subjects' data in detail.

\section{Discussion \& Conclusions}

HMMs can provide various outcomes that are useful for clinical research on observational data.
In particular, HMMs reveal latent states and their transitions from a set of observed variables in an unsupervised manner.
It is often difficult to understand the complex relationship between the states and observed variables.
\toolname explains the relationship between latent states and observed variables transparently.
Visual summary of states and their transition remind researchers of what the outputs are with the observed variables, which are familiar to them in the disease domain.
Showing a summary of relationship between inputs and outputs can help researchers to understand the models in context of their clinical domains.

Clinical researchers build and test hypotheses iteratively by comparing different subject cohorts.
The cohorts are often challenging to describe when using sophisticated criteria for grouping patients;
\toolname provides various ways to build cohorts using all views and allows direct construction and refinement of queries using \query, setting intuitive filtering actions on any views.
The flexible methods for constructing filters was useful for clinical experts to formulate and test their hypotheses.
The features empower researchers to test their hypotheses without leaving the tool because they construct cohorts and compare the patterns out of those cohorts visually.
\cohort enables users to maintain and share such cohorts with colleagues as they iteratively refine their analysis.
Using these features, researchers generated clinically useful insights for potential future publication in clinical journals~\cite{dpvis}.

\toolname provides an overview and details of individual subjects' records in a single application.
Summaries are helpful, and details are, too.
Multiple, coordinated, and interactive visualizations integrated into a single system like \toolname make it possible to see details without losing the context of overview.
Clinical researchers have extensive knowledge about individual subjects, which may not be captured in the dataset.
Visualizations in various abstraction levels can help clinical researchers make sense of patterns in light of their domain knowledge more easily.
It is especially useful to translate experts' feedback into model iterations.
For instance, \toolname helps users to estimate the usefulness of the model instance and helps them make decisions whether they should adjust the number of states for next iterations.
To achieve the goal, visual analytics applications for clinical research need to provide multiple levels of information.

There is great potential in application of AI and machine learning (ML) algorithms to discover hidden patterns and new knowledge out of longitudinal health data.
It can help clinicians to make breakthrough in the field of medicine.
Yet, the challenges are to make such AI and ML algorithms useful for their research.
To leverage clinicians' domain expertise fully, the system needs to continuously tap into their expert knowledge.
This demo shows the usefulness of interactive visualizations in human-centered AI for clinical research by making patterns more transparent and by allowing them researchers to interactively test their hypotheses.
How can we extend this approach for other scenarios and tasks (e.g., predictive modeling) in data analysis?
There's no single right answer to this question.
It is necessary for clinicians, ML researchers, and visualization researchers to collaboratively explore the uncharted areas.

\section*{Acknowledgment}
We wish to thank the T1DI Study Group for their help in this work.
This work was supported in part by JDRF (1-IND-2019-717-I-X, 1-SRA-2019-722-I-X, 1-SRA-2019-723-I-X, 1-SRA-2019-719-I-X, 1-SRA-2019-721-I-X, 1-SRA-2019-720-I-X).

\bibliographystyle{ACM-Reference-Format}
\bibliography{dpvis}

\end{document}